\documentclass[preprint,12pt, a4paper]{elsarticle}

\usepackage{amsmath,amsbsy}

\usepackage{subfig}

\usepackage{hyperref}

\biboptions{round, authoryear}

\usepackage{threeparttable}

\usepackage{multirow}

\begin{document}

	\begin{frontmatter}
	
		\title{Recreating the OSIRIS-REx Slingshot Manoeuvre from a Network of Ground-Based Sensors}
        
		\author[1]{Trent Jansen-Sturgeon}
		\author[1]{Benjamin A. D. Hartig}
		\author[2]{Gregory J. Madsen}
		\author[1]{Philip A. Bland}
		\author[1]{Eleanor K. Sansom}
		\author[1]{Hadrien A. R. Devillepoix}
		\author[1]{Robert M. Howie}
		\author[1]{Martin Cup\'ak} 
		\author[1]{Martin C. Towner}
		\author[1]{Morgan A. Cox}
		\author[1]{Nicole D. Nevill} 
		\author[1]{Zacchary N. P. Hoskins} 
		
		\author[3]{Geoffrey P. Bonning} 
		\author[4]{Josh Calcino} 
		\author[5]{Jake T. Clark} 
		\author[6]{Bryce M. Henson} 
		\author[8]{Andrew Langendam} 
		\author[9]{Samuel J. Matthews} 
		\author[10]{Terence P. McClafferty} 
		\author[8]{Jennifer T. Mitchell} 
		\author[11]{Craig J. O'Neill} 
		\author[7]{Luke T. Smith} 
		\author[8, 12]{Alastair W. Tait} 

		\address[1]{School of Earth and Planetary Sciences, Curtin University, GPO Box U1987, Perth, WA, 6845, Australia}
		\address[2]{Lockheed Martin Australia, Barton, ACT, 2600, Australia}
		
		\address[3]{Research School of Earth Sciences, Australian National University, Canberra, ACT, 2601, Australia}
		\address[4]{School of Mathematics and Physics, The University of Queensland, QLD, 4072, Australia}
		\address[5]{Centre for Astrophysics, University of Southern Queensland, Toowoomba, QLD, 4350, Australia}
		\address[6]{Laser Physics Centre, Research School of Physics and Engineering, The Australian National University, Canberra, ACT, 2601, Australia}
		\address[7]{Department of Earth and Planetary Sciences, Macquarie University, Sydney, NSW, 2109, Australia}
		\address[8]{School of Earth, Atmosphere and Environment, Monash University, Clayton, Victoria, 3800, Australia}
		\address[9]{Geological Survey of New South Wales, NSW, Australia}
		\address[10]{College of Education, Charles Darwin University, Darwin, NT, 0909, Australia}
		\address[11]{Macquarie Planetary Research Centre, Macquarie University, Sydney, NSW, 2109, Australia}
        \address[12]{Biological and Environmental Sciences, University of Stirling, Stirling, Scotland, UK}

		\begin{abstract}
    		
            Optical tracking systems typically trade-off between astrometric precision and field-of-view. In this work, we showcase a networked approach to optical tracking using very wide field-of-view imagers that have relatively low astrometric precision on the scheduled OSIRIS-REx slingshot manoeuvre around Earth on September 22nd, 2017. As part of a trajectory designed to get OSIRIS-REx to NEO 101955 Bennu, this flyby event was viewed from 13 remote sensors spread across Australia and New Zealand to promote triangulatable observations. Each observatory in this portable network was constructed to be as lightweight and portable as possible, with hardware based off the successful design of the Desert Fireball Network.
            
    		Over a 4 hour collection window, we gathered 15,439 images of the night sky in the predicted direction of the OSIRIS-REx spacecraft. Using a specially developed streak detection and orbit determination data pipeline, we detected 2,090 line-of-sight observations. Our fitted orbit was determined to be within about 10~km of orbital telemetry along the observed 109,262~km length of OSIRIS-REx trajectory, and thus demonstrating the impressive capability of a networked approach to SSA.
    		
		    
		\end{abstract}
		
		\begin{keyword}
			OSIRIS-REx \sep Networked SSA \sep Desert Fireball Network \sep Streak Detection \sep Triangulation \sep Orbit Determination \sep Telemetry Comparison \sep FireOPAL
		\end{keyword}
		
	\end{frontmatter}


\pagebreak
\section{Introduction}
Ever since the first Near Earth Object (NEO) was discovered in 1898 \citep{miller_determination_2002}, humanity has been concerned of the threat posed by these objects. A variety of methods have been developed to determine their orbits. Traditionally, NEO orbits have been established using angles-only measurements from a single observation point. These methods include Gauss's method \citep{gauss_theory_1857}, Laplace's method \citep{klokacheva_determination_1991}, Gooding's method \citep{gooding_new_1997}, and the Double-R algorithm \citep{der_new_2012}. However, these single-viewpoint techniques tend to perform better with measurements spanning a large time period.

Modern methods have been developed for the obit determination of NEO's using temporally close measurements that span short-arcs, such as the admissible region approach of \citet{milani_astrometry_2005} and \citet{tommei_orbit_2007}. However these methods determine a set of physically acceptable orbits and are still prone to considerable error. An alternative approach to the short-arc problem uses two observatories taking simultaneous measurements to triangulate the NEO, which results in greater accuracy over a shorter time window when compared to the single observatory approach, even when data density is matched. \citet{eggl_refinement_2011} describes a method of NEO triangulation using measurements from two heliocentric satellites, and \citet{eggl_prospects_2014} explains the same concept between the Gaia satellite and ground-based observations.

Triangulation has also been used by fireball networks since 1959 to track and dynamically model small NEO's from meteor and fireball observations from multiple distributed sensors as they ablate through our upper atmosphere. Of the 60,000 meteorites in collections around the world, only about 32 have known orbits as of 2018 \citep{granvik_identification_2018}. Five of these have been discovered using the Desert Fireball Network (DFN); the largest fireball network in the world. 

With fireball networks as inspiration, we propose using the same distributed multi-observatory approach to triangulate objects beyond our atmosphere. The question is whether it is possible to accurately determine the orbit of a heliocentric object using many relatively low-resolution ground-based sensors.

To answer this and test the validity of such an approach, we observed the OSIRIS-REx spacecraft's flyby manoeuvre on September 22nd, 2017, using Australasia as our triangulation baseline. Additionally, to verify the accuracy of such a method, we recreate its NEO-like orbit and compare it to well known orbital telemetry. 

\section{Observatory Hardware}\label{sec:hardware}
\begin{figure}
    \centering
    \includegraphics[width=0.8\textwidth]{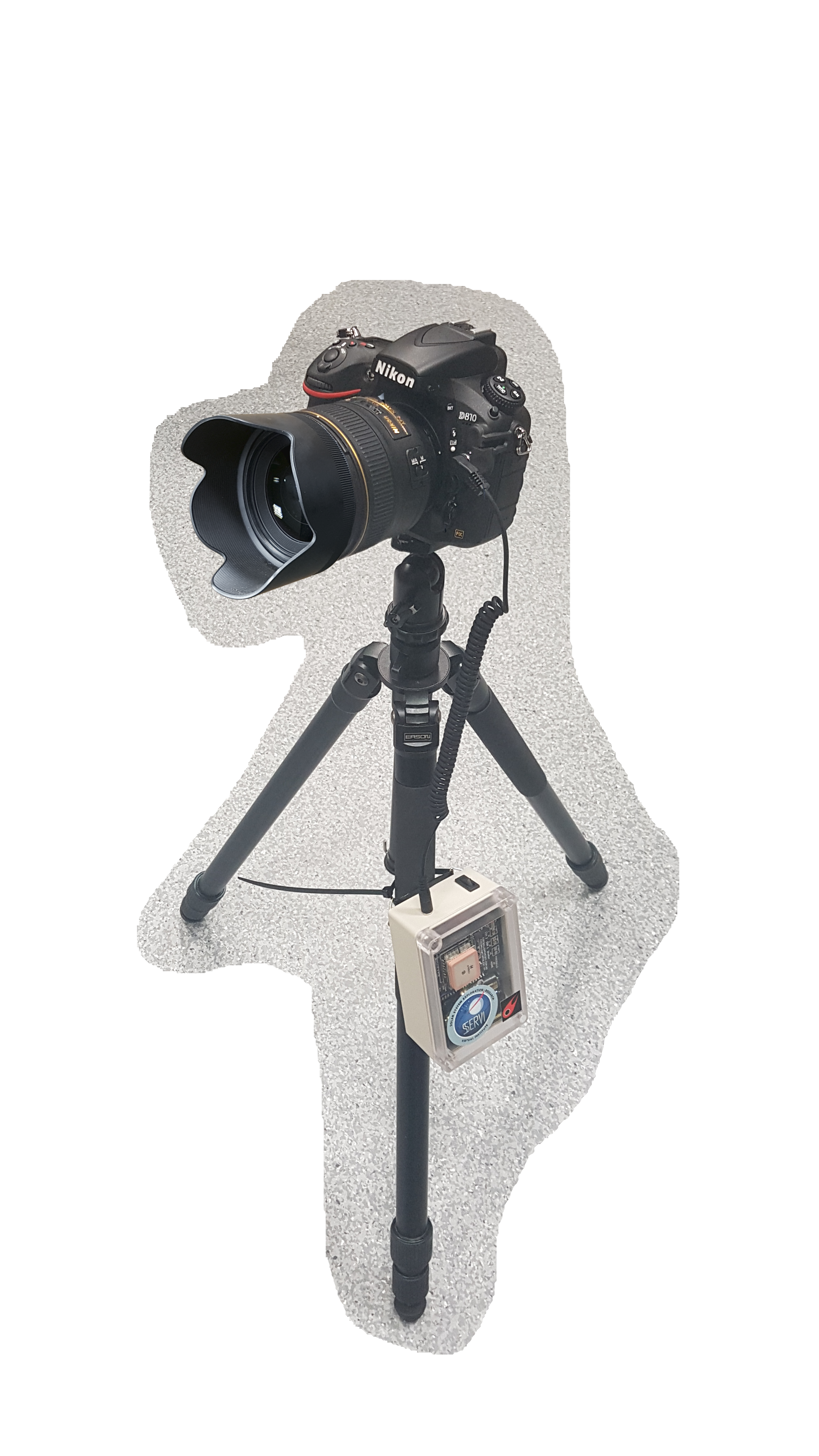}
    \caption{One of the observatories used in the OSIRIS-REx observation campaign, complete with the custom designed triggering unit.}
    \label{fig:orex_obs}
\end{figure}

Rapidly deployable proof-of-concept observatories were developed to demonstrate a networked approach to SSA by imaging the OSIRIS-REx spacecraft during its flyby manoeuvre. As the chosen observation locations were spread out to very different places across the country, these observatories were constructed to be portable, light-weight, and easy to setup and operate in the field by non-specialists. The design consisted of only the bare-bones components and could be separated into two parts - a small custom designed triggering unit and the imager itself - all able to be transported in a small case as carry on luggage as shown in Fig.~\ref{fig:orex_obs}.

The triggering unit is a key piece of hardware that takes advantage of the DFN's existing hardware \citep{howie_how_2017}. It contains a micro-controller linked to a Global Navigation Satellite System (GNSS) chip to activate the camera's shutter at predetermined times. The triggering unit also contains an SD card to record the imager's geodetic coordinates and absolute time information for later data processing. The predetermined times were synchronised across the continental-scale network at a cadence of 6 seconds, triggered to sub-millisecond precision. This synchronisation between observatories allowed for individual pointwise triangulation and possible reconstruction of spacecraft geometry through light curve analysis.

The imager consists of a Nikon D810 36 megapixel DSLR full frame sensor paired with either a Sigma Art 85\,mm f/1.4 (11 observatories) or a Nikon Nikkor 105\,mm f/1.4 lens (2 observatories). The Nikon camera was set to 3200 ISO, maximum aperture (f/1.4), and 4 second exposure time.

The majority of these portable observatories provide a 24\,x\,16~degree field-of-view or 385~square degrees, resulting in a resolution of about 11.5~arcsec per pixel. This would be classed as relatively low-resolution when compared to a typical SSA telescope, such as the Falcon Telescopes \citep{chun_new_2018}. Sponsored by the United States Air Force Academy (USAFA), the Falcon Telescopes support a 0.5~meter, f/8.1 lens and 0.65~arcsec per pixel, equating to a coverage of only about 12~square degrees. While these typical SSA telescopes are fantastic at refining the orbit of an already catalogued space object, they lack the field-of-view necessary for blind target acquisition and even have trouble capturing targets of high uncertainty, which is where large field-of-view sensors excel. 


\section{OSIRIS-REx Observation Campaign} 
The site selection and observatory pointing for the observation campaign were both based around the OSIRIS-REx slingshot orbit as predicted by NASA. This predicted orbit was gathered from the NASA Horizons web service (\url{https://ssd.jpl.nasa.gov/horizons.cgi}) on August 17th, 2017; about a month before the flyby manoeuvre. 

\subsection{Site Selection}
The optimal triangulation baseline for any target is when observations from two or more observatories meet at 90 degrees. Due to the large range of OSIRIS-REx even at closest approach, this baseline is greater than the diameter of the Earth. Therefore, to best demonstrate the distributed network approach, observatory sites were chosen to maximise the observation baseline while avoiding light polluted areas and adverse weather conditions on the night of the 22nd of September, 2017. As such, 12 sites across Australasia were chosen in advance with a few alterations very last minute depending on the weather forecast. 

\begin{figure}
    \centering
    \includegraphics[width=\textwidth]{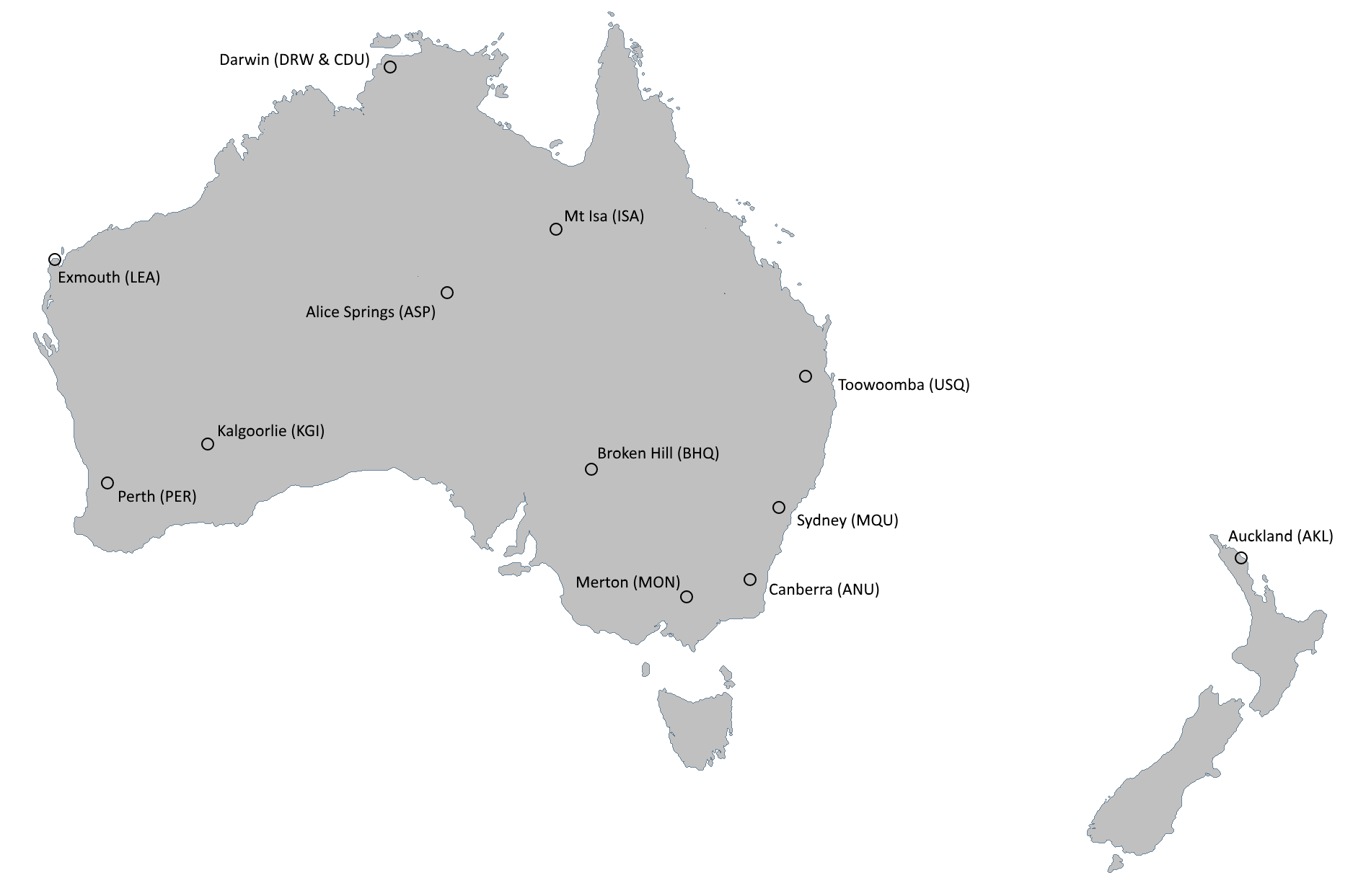}
    \caption{Sensor locations on the night of the OSIRIS-REx slingshot manoeuvre.}
    \label{fig:obs_locations}
\end{figure}

The portable observatories on these 12 sites were operated by teams from 6 different Australian institutions. A 13th observatory was setup alongside the Darwin node to allow direct performance comparisons between the 85\,mm and 105\,mm lenses. The final viewing sites for the OSIRIS-REx flyby are shown in Fig.~\ref{fig:obs_locations} and specified in Table~\ref{tab:locations}.

\subsection{Pointing}
The OSIRIS-REx slingshot trajectory spanned a large fraction of the sky, considerably more than the field-of-view of our observatories. Therefore, a number of manual re-points were required by the observatory operators over the collection window. This collection window started from about 2017-09-22 12:35 UTC, when OSIRIS-REx was predicted to be at our sensor's limiting brightness of 12th magnitude, and finished just before it fell below the horizon over 4 hours later. Fortunately the Moon was set for the entire collection period, giving us the best chance at capturing the faint reflection of the OSIRIS-REx spacecraft.

To avoid frequent re-pointing while still keeping the OSIRIS-REx spacecraft near the centre of the field-of-view, the predicted altitude/azimuth pointing information was segmented into 5 degree arcs from each observatory's perspective. The centroid of these arc segments dictated where the sensor was to be pointed and at what time. This resulted in about 25 re-points over the course of 4 hours, with an initial spacing of 20 min down to about 4 min by the end as OSIRIS-REx increased in apparent angular velocity as it approached the horizon. 

The actual process of pointing the sensors on the night was rough due to the imprecise nature of the compass (corrected for magnetic declination) and an inclinometer used to aim the lens. However, the 5 degree arc segment was designed to be sufficiently smaller than the field-of-view of the lens to cater for this known inaccuracy. This rudimentary pointing method allowed less experienced operators to successfully observe the target, without the need of sky charts or even the need to carry a computer.

\subsection{Gathered Data}
Over the course of the observation campaign, 15,439 images were taken of the night sky from 12 different viewpoints. Observer location and absolute time records were also gathered alongside these images at the moment of shutter actuation using the custom triggering box described in Section~\ref{sec:hardware}. The locations and collected data for each observatory is summarised in Table~\ref{tab:locations}. 

\begin{table}[h!]
    \centering
    \caption{The final locations of the portable observatories on the night of September 22nd, 2017. The OSIRIS-REx viewing window and number of images captured are also shown for each site. Each observatory supported a 85mm lens besides two locations (*), which used the 105mm lens.}
    \begin{tabular}{|l|cc|cc|}
        \hline
        \multirow{2}{*}{Location (Codename)} & \multirow{2}{*}{Latitude} & \multirow{2}{*}{Longitude} & Observed & Collected \\
         &  &  & Window & Images \\
        \hline
        Auckland (AKL)      & -35.548$^{\circ}$ & 174.303$^{\circ}$ & 1h 36m &   546 \\ 
        Canberra (ANU)      & -36.522$^{\circ}$ & 149.228$^{\circ}$ & 4h 06m & 2,147 \\ 
        Alice Springs (ASP) & -23.683$^{\circ}$ & 133.928$^{\circ}$ & 3h 32m & 1,586 \\ 
        Broken Hill (BHQ)*  & -31.847$^{\circ}$ & 141.203$^{\circ}$ & 4h 15m & 1,564 \\ 
        Darwin (CDU)        & -13.044$^{\circ}$ & 130.995$^{\circ}$ & 3h 35m & 1,398 \\ 
        Darwin (DRW)*       & -13.044$^{\circ}$ & 130.995$^{\circ}$ & 3h 31m & 1,415 \\ 
        Mount Isa (ISA)     & -20.903$^{\circ}$ & 139.440$^{\circ}$ & 3h 38m & 1,535 \\ 
        Kalgoorlie (KGI)    & -30.751$^{\circ}$ & 121.760$^{\circ}$ & 3h 23m & 2,031 \\ 
        Learmonth (LEA)     & -22.401$^{\circ}$ & 114.039$^{\circ}$ & 2h 53m & 1,050 \\ 
        Melbourne (MON)     & -36.975$^{\circ}$ & 145.706$^{\circ}$ & 1h 17m &   719 \\ 
        Sydney (MQU)        & -33.536$^{\circ}$ & 151.295$^{\circ}$ & 2h 30m &    NA \\ 
        Perth (PER)         & -32.406$^{\circ}$ & 116.725$^{\circ}$ & 0h 18m &   139 \\ 
        Toowoomba (USQ)     & -27.825$^{\circ}$ & 152.101$^{\circ}$ & 3h 25m & 1,309 \\ 
        \hline
    \end{tabular}
    \label{tab:locations}
\end{table}

This is far too much data to be reduced manually. In order to recreate the OSIRIS-REx slingshot orbit from the gathered imagery, an automated data reduction pipeline was required.

\section{Recreating the OSIRIS-REx Orbit}
To construct the flyby orbit of the OSIRIS-REx satellite, we must first detect and extract the start and end points of the recorded spacecraft's streak from the mass of images captured from the various observation sites in a process called streak detection. Additionally, if OSIRIS-REx is visible, light curves can be determined throughout the manoeuvre from different points-of-view, leading to possible shape reconstruction. 

Using the extracted streak information from the images, we determine the most likely orbit of the OSIRIS-REx spacecraft given the mass of measurements. This recreated orbit fit is then compared to both NASA Horizons predicted path and actual OSIRIS-REx telemetry of the flyby trajectory.

\subsection{Streak Detection}\label{ssec:streak_detection}
\begin{figure}
    \centering
    \includegraphics[width=\textwidth]{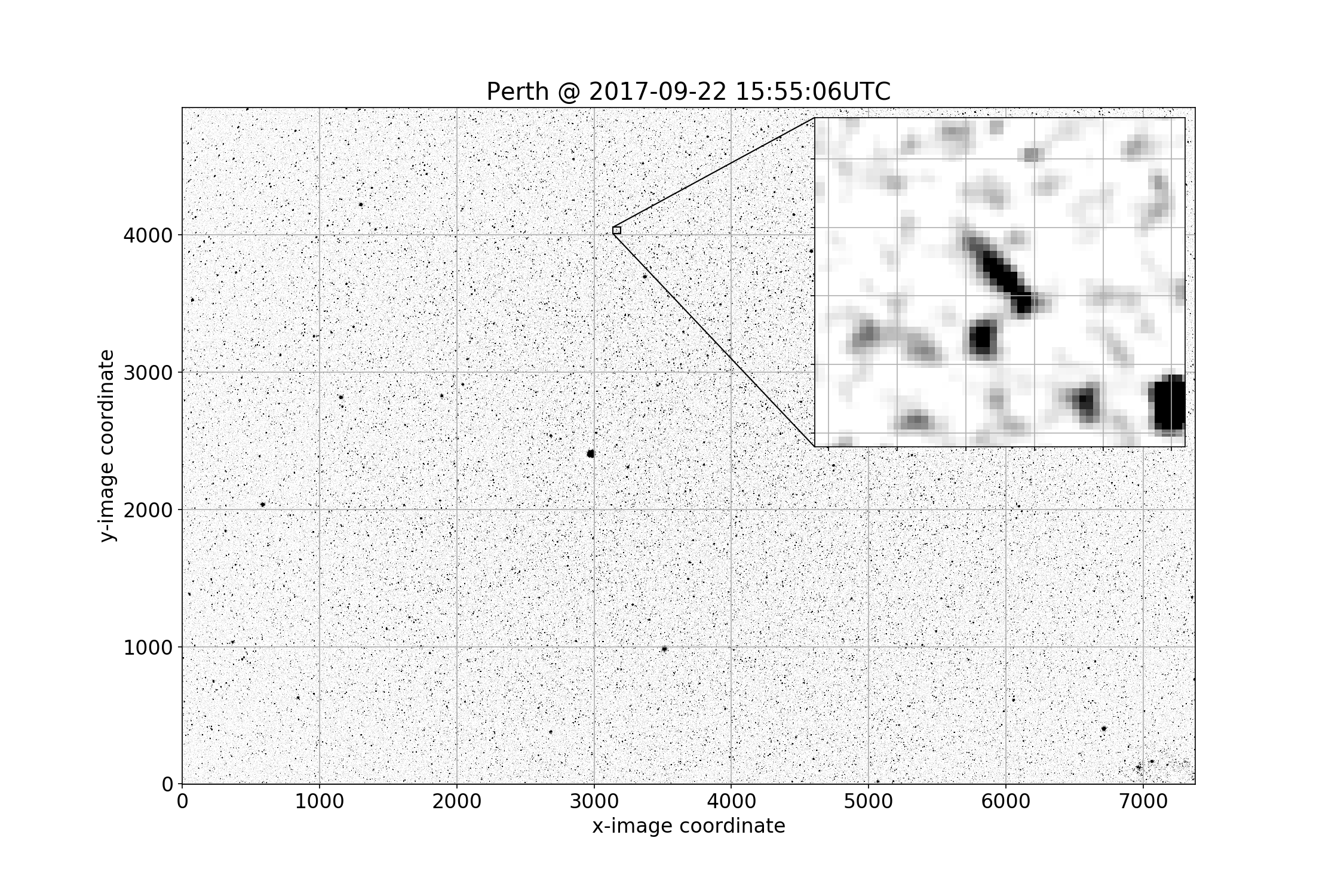}
    \caption{An example of a full-framed image with an enlarged region to show the scale of an OSIRIS-REx streak. For reference, this image is $24^{\circ} \times 16^{\circ}$.}
    \label{fig:streak_example}
\end{figure}

Of the 15,439 captured images, the OSIRIS-REx spacecraft was detected in 1,045 of them from 9 viewpoints. Unfortunately, the data collected from 4 of the locations were unusable due to technical or weather related issues. Additionally, out of the sites that did capture valuable imagery, OSIRIS-REx was not observed in all of them because of either human pointing error, unfavourable lighting, or OSIRIS-REx was simply too faint to detect. For streak scale, Fig.~\ref{fig:streak_example} shows an OSIRIS-REx streak detection in a full-framed image.

\begin{figure}
    \centering
    \includegraphics[width=\textwidth]{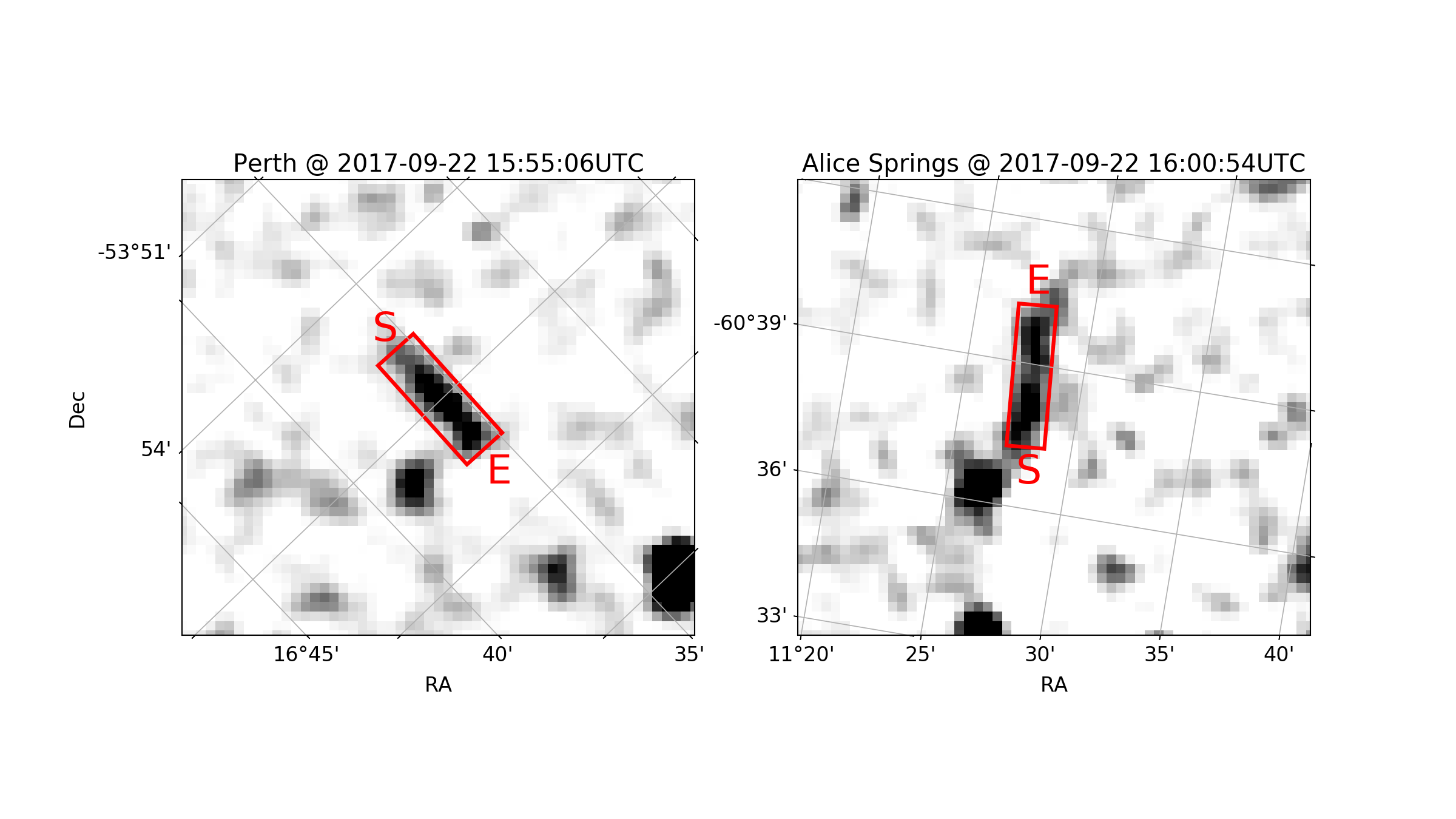}
    \caption{Two thumbnail examples of OSIRIS-REx within 4 second long-exposures as viewed from Perth and Alice Springs, respectively. The OSIRIS-REx streak has been highlighted and its direction of travel indicated, where "S" is the start and "E" is the end of the exposure. The full-frame image had been astrometrically calibrated before thumbnail creation.}
    \label{fig:detected_streaks}
\end{figure}

To autonomously extract the OSIRIS-REx streak information from the images, we designed a sensitive streak detection algorithm. Firstly, adjacent images captured from the same sensor are astrometrically aligned, smoothed and subtracted from one-another to highlight any differences. The smoothly varying background sky is then removed before any statistically significant streaks are identified. Thumbnails centred around these streaks are saved, with the encompassed stars used to astrometrically and photometrically calibrate the target streaks. Two examples of OSIRIS-REx thumbnails are shown in Fig.~\ref{fig:detected_streaks}, with the streak itself highlighted.

By tuning the streak detection algorithm towards sensitivity, we introduce a significant number of noise-level artefacts. However, most of these artefacts can be discarded as they do not resemble a satellite streak or they are not in the predicted region of the image to be OSIRIS-REx. The OSIRIS-REx search region within each images is large enough to cater for any inaccuracies in the predicted trajectory, but small enough to avoid most of the aforementioned artefacts. 

Generally, a streak's direction of travel can be identified by comparing it to past and future images from the same sensor. However, sometimes streaks do not appear in the adjacent images, making the streak's orientation ambiguous. Due to the faintness of OSIRIS-REx, the streak detection algorithm detected a significant number ambiguous streaks. This ambiguity is later resolved at the orbit determination stage. 

\subsection{Light Curves}
\begin{figure}
    \centering
    \includegraphics[width=\textwidth]{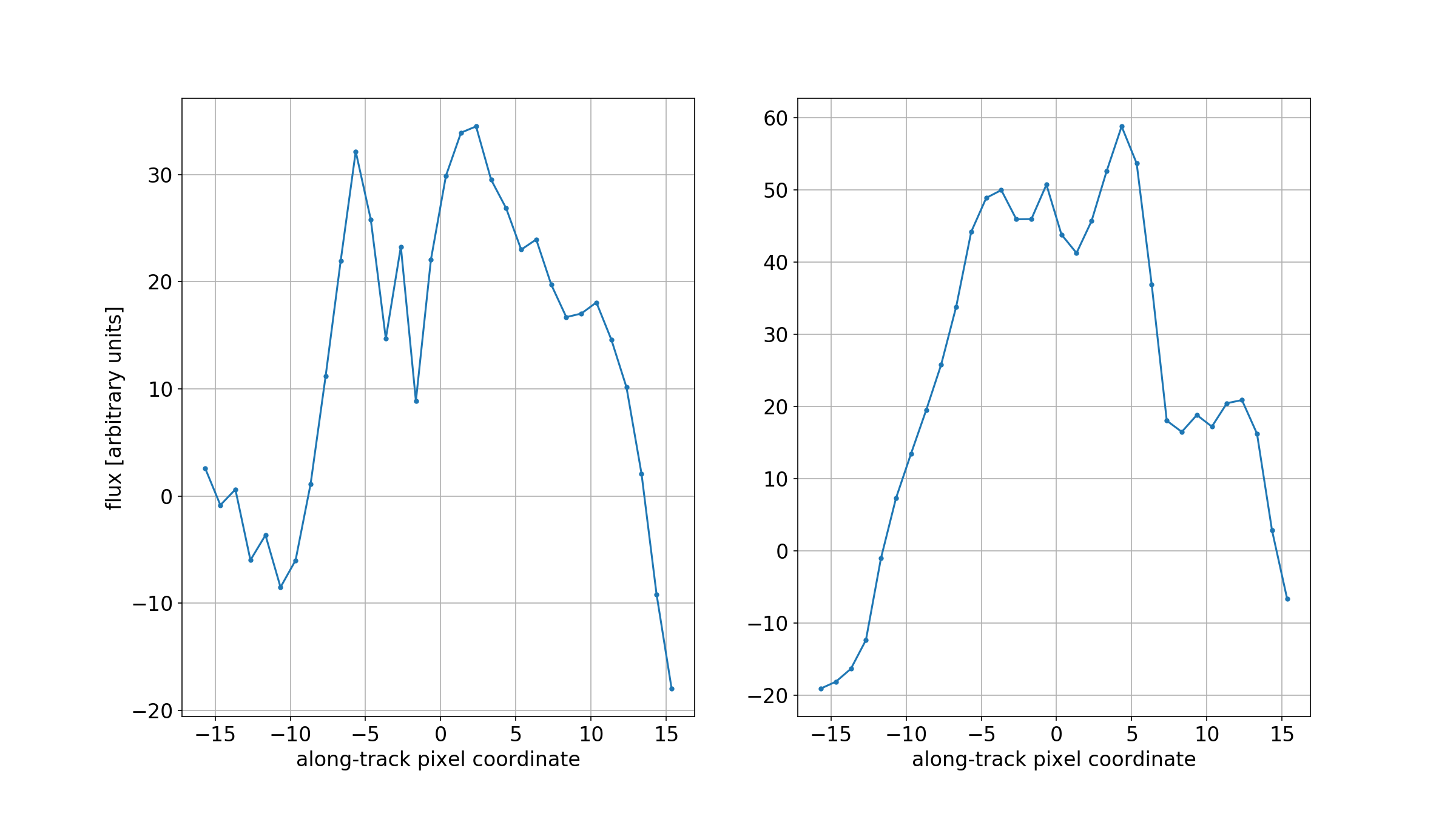}
    \caption{The flux measured along the OSIRIS-REx streaks corresponding to those highlighted in Fig.~\ref{fig:detected_streaks}, where the zeroth pixel coordinate coincides to the middle of the streak.}
    \label{fig:some_light_curves}
\end{figure}

\begin{figure}
    \centering
    \includegraphics[width=\textwidth]{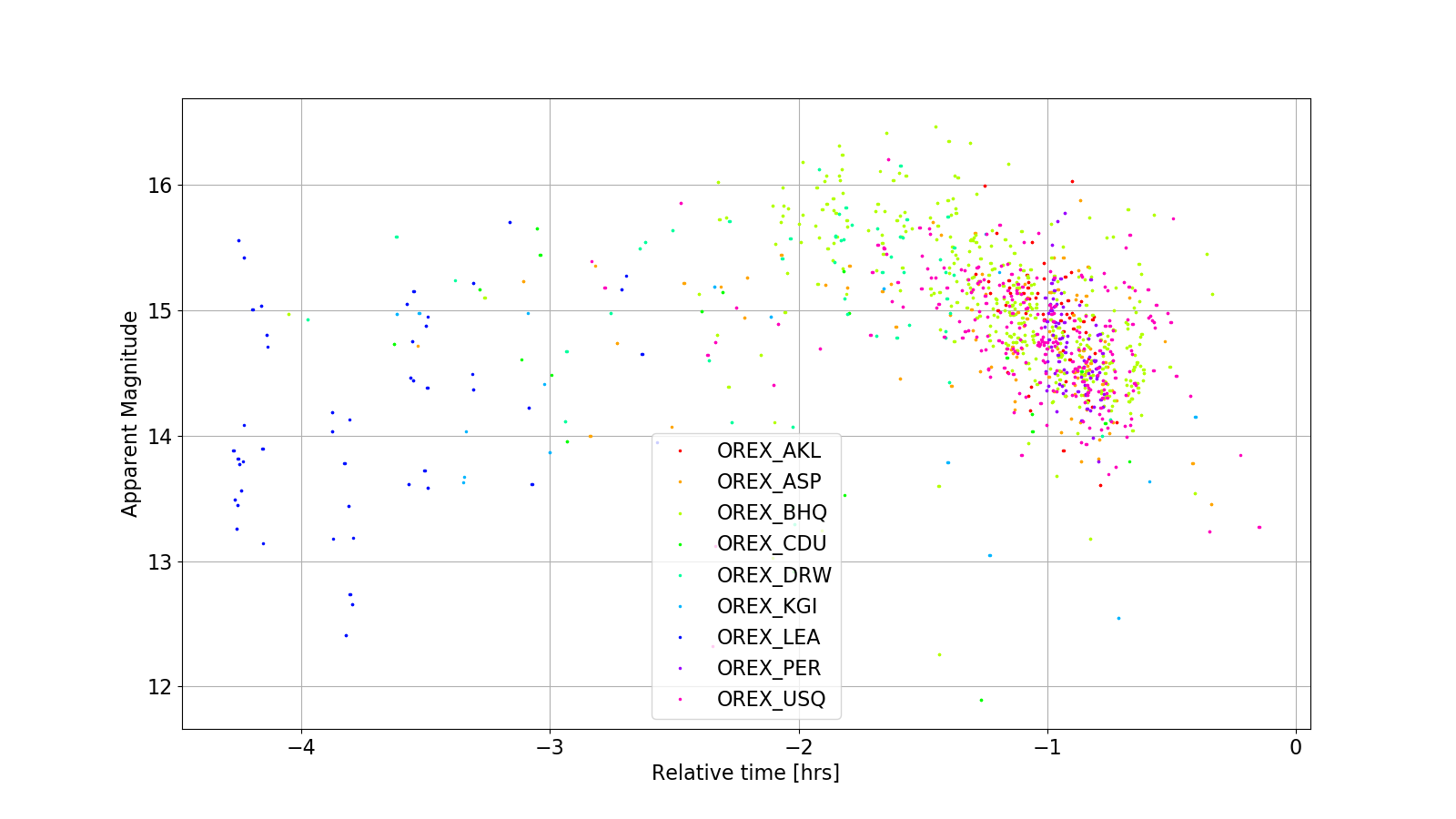}
    \caption{Apparent magnitude of the OSIRIS-REx spacecraft from different observation locations throughout the observation campaign. Times are relative to the closest approach; 2017-09-22 16:51:50.818 UTC. See Fig.~\ref{fig:obs_locations} for the specific sensor locations.}
    \label{fig:all_light_curves}
\end{figure}

After a streak has been identified and measured as described in Section~\ref{ssec:streak_detection}, the light curves are then determined by examining the pixel brightness along the length of the streak. The pixel brightness is converted into apparent flux along the streak using the photometric calibration obtained in the streak detection step. These simultaneous light curves gathered from various synchronised viewpoints would allow for the complete reconstruction of the OSIRIS-REx spacecraft's geometry. Fig.~\ref{fig:some_light_curves} shows how the brightness changes within a couple of example streaks and Fig.~\ref{fig:all_light_curves} shows OSIRIS-REx's change in apparent magnitude over the flyby viewing window from all perspectives.

Perhaps surprisingly, the apparent magnitude of the OSIRIS-REx spacecraft begins relatively bright, becoming its dimmest about two hours before the closest approach. One might think that the brightness of the spacecraft should increase monotonically as the observation range decreases. This would be true if all other factors were ignored, such as the satellite's orientation, irregular shape, varying albedo, and in particular phase angle. The phase angle of OSIRIS-REx throughout the observation campaign went from 15 degrees to over 90 degrees, where zero degrees corresponds to maximum solar reflectance. 

Additionally, the spread of these apparent magnitudes as shown in Fig.~\ref{fig:all_light_curves} can be due to a multitude of reasons, including Sun glint off the solar panels, complex geometry of the satellite, and locally varying atmospheric conditions.

\subsection{Orbit Determination}\label{ssec:orbit_determination}
Calculating an orbit given all the line-of-sight measurements is merely an optimisation/fitting problem. To determine this best fit orbit, we choose to use a batch least squares approach that iteratively adjusts an initial orbit guess (comprising of six orbital parameters) to minimise the sum of the squared residuals, where in our case, a residual is the angular difference between the observed line-of-sight and the predicted line-of-sight at the time of the observation. The predicted lines-of-sight are calculated from the initial orbit guess using a two-body Earth-centred (J2000) hyperbolic orbit propagator due to the nature of the flyby event.

This least squares minimisation procedure is repeated until a local minimum is reached as determined by the residual Jacobian. If any residuals are outside three median-absolute-deviations at the point of this local minimum, they are classed as outliers and are removed one-by-one before the batch least squares procedure is re-run on the remaining data. An outlier could unavoidably arise during the image differencing step within the streak detection algorithm if an observed OSIRIS-REx streak occultates a star, for example. This outlier rejection step is repeated after every least squares operation until no outliers remain to contribute to the optimal orbit fit.

In late 2018 at the AMOS conference, a preliminary orbit fit was presented using the above procedure that merely used the centroids of the streaks \citep{jansen-sturgeon_fireopal:_2018} and therefore did not require any ambiguous streak handling as mentioned in Section~\ref{ssec:streak_detection}. In order to include all avaliable line-of-sight measurements, the orbit determination algorithm was later refined to use the start and end pointing coordinates of the OSIRIS-REx streaks. 

To combat the streak ambiguity problem, the ambiguous streak's direction of travel is determined before any optimisation techniques are applied by comparing to the known NASA Horizons predicted orbit. Any ambiguous streaks that appear to be heading in the wrong direction are flipped; i.e. if their dot product with the projected velocity vector at that time is less than zero. This streak flipping process ultimately aids in the accuracy of the overall orbit fit.

\begin{figure}
    \centering
    \includegraphics[width=\textwidth]{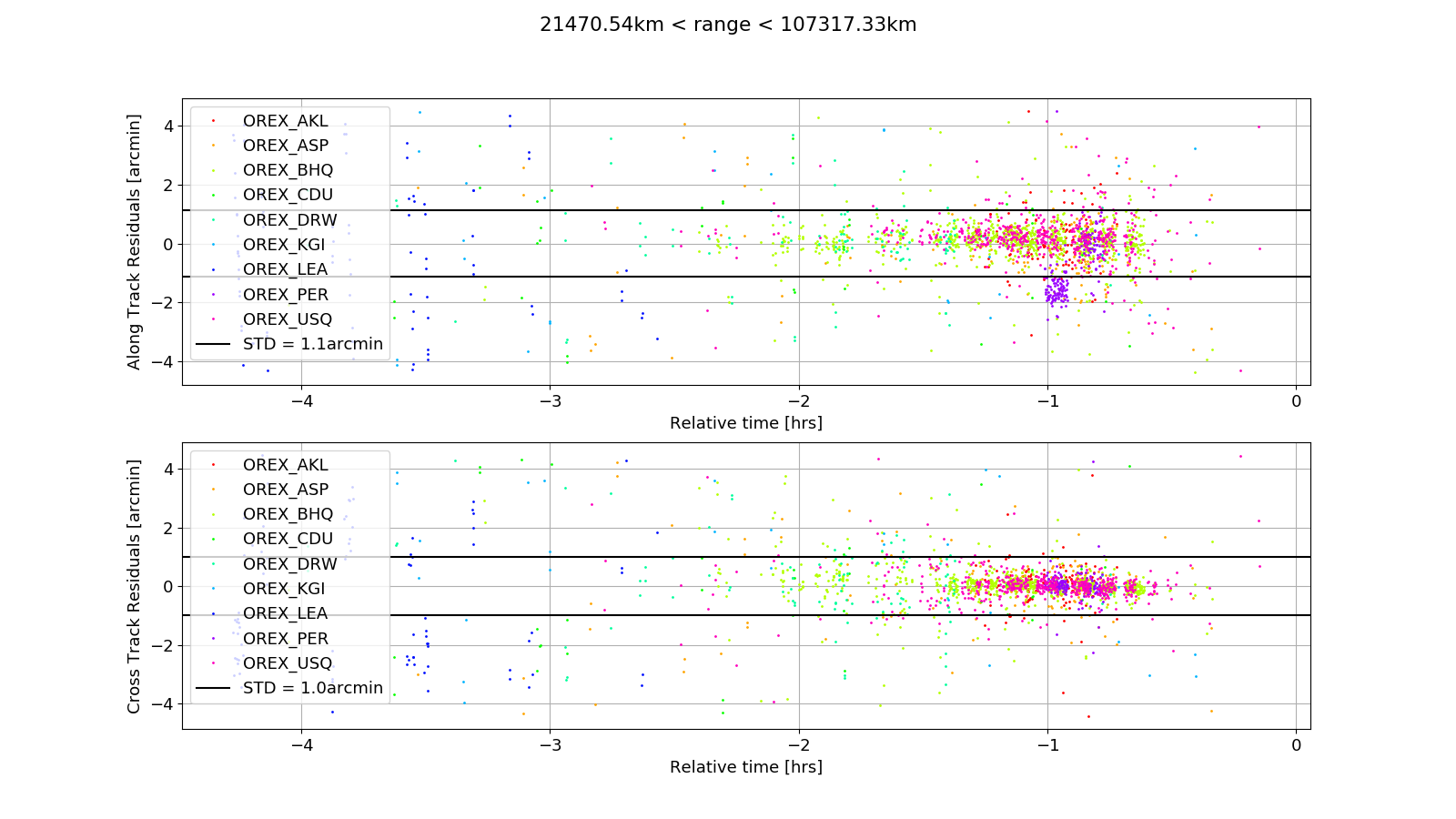}
    \caption{The angular difference between the observed line-of-sight measurements and the predicted lines-of-sight over time given the fitted hyperbolic orbit, otherwise known as residuals. The sensors are colour coded and the residual standard deviation is given in black. The time is relative to the point of closest approach; 2017-09-22T16:51:50.818UTC.}
    \label{fig:residuals}
\end{figure}

The adjusted orbit determination code corrected 198 of the 353 ambiguous streaks that were found to be oriented the wrong way. Additionally, we identified 99 out of the 2,090 measurements as outliers, which were discarded from the final orbit fit. The remaining measurements are represented in Fig.~\ref{fig:residuals} as the orbit residuals, and the final calculated hyperbolic orbital elements are listed in Table~\ref{table:orbits}.

To validate the use of this simple two-body approximation, we calculate the deviation from the strongest perturbation (namely J2) over our 4 hour observation window to be about 30 meters. This is far less than the error on the perigee alone, see Table~\ref{table:orbits}, and therefore would negligibly affect our overall orbit solution.



\subsection{Orbit Comparison}\label{ssec:orbit_comparision}
\begin{figure}
    \centering
    \includegraphics[width=\textwidth]{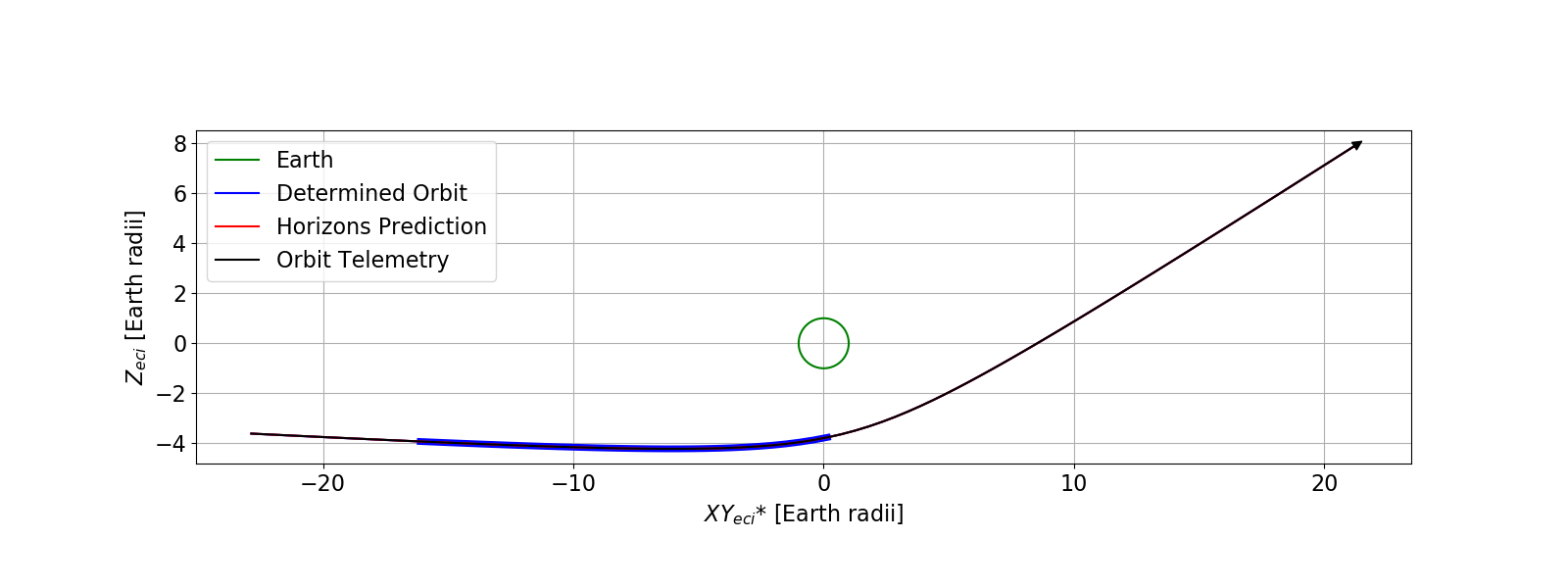}
    \caption{The three hyperbolic orbits from Table~\ref{table:orbits} compared in the Earth Centered Inertial (GCRS) frame, spanning 6 hours either side of the closest approach. The orbits are so similar that they appear as one orbit in this plot. To visualise the subtle orbit differences, please refer to Fig.~\ref{fig:cross-tracks}. The thicker blue trajectory is highlighting the 4 hour long observed section of the determined orbit.
    *The $XY_{eci}$ axis is angled at an azimuth equal to the longitude of ascending node of the determined orbit as to best represent the hyperbolic orbits in a 2D plane.}
    \label{fig:orbits}
\end{figure}

Now that we have calculated OSIRIS-REx's orbital trajectory, we have an opportunity to compare this against the original predicted orbit and actual OSIRIS-REx telemetry. As discussed, the predicted orbit was gathered from NASA Horizons web service (\url{https://ssd.jpl.nasa.gov/horizons.cgi}) on August 17th, 2017, about a month before the flyby event. The orbital telemetry was collected from the same location about a year after the event, and has not been altered since. Table~\ref{table:orbits} compares the hyperbolic orbital elements of the determined orbit to the prediction and telemetry orbits, while Fig.~\ref{fig:orbits} compares them all visually.

\begin{table}[h!]
    \centering
    \caption{The hyperbolic orbital elements at the time of closest approach, 2017-09-22 16:51:50.818 UTC, expressed in the Earth Centred Inertial (GCRS) coordinate frame. In order, p, e, i, $\omega$, $\Omega$, and M correspond to the perigee, eccentricity, inclination, argument of perigee, longitude of the ascending node, and the mean anomaly respectively.}
    \begin{threeparttable}
    \begin{tabular}{ |l||c|c|c| }
        \hline
        Orbital  & Horizons            & Determined & Orbit              \\ 
        Elements & Prediction\tnote{a} & Orbit      & Telemetry\tnote{b} \\ 
        \hline
        p [km]         & 23595.52402  & 23590   $\pm$ 8      & 23591.78261  \\
        e              & 3.29654      & 3.2971  $\pm$ 0.0006 & 3.29620      \\ 
        i [deg]        & 84.84226     & 84.831  $\pm$ 0.001  & 84.83158     \\ 
        $\omega$ [deg] & 284.52092    & 284.528 $\pm$ 0.004  & 284.52307    \\ 
        $\Omega$ [deg] & 185.61554    & 185.606 $\pm$ 0.001  & 185.61538    \\ 
        M [deg]        & 0.18841      & 0.180   $\pm$ 0.004  & 0.15535      \\
        \hline
    \end{tabular}
    \begin{tablenotes}\begin{footnotesize}
        \item[a] Gathered from NASA Horizons Web-Interface on 2017-08-17.
        \item[b] Gathered from NASA Horizons Web-Interface post slingshot.
    \end{footnotesize}\end{tablenotes}
    \end{threeparttable}
    \label{table:orbits}
\end{table}

\begin{figure}
    \centering
    \includegraphics[width=\textwidth]{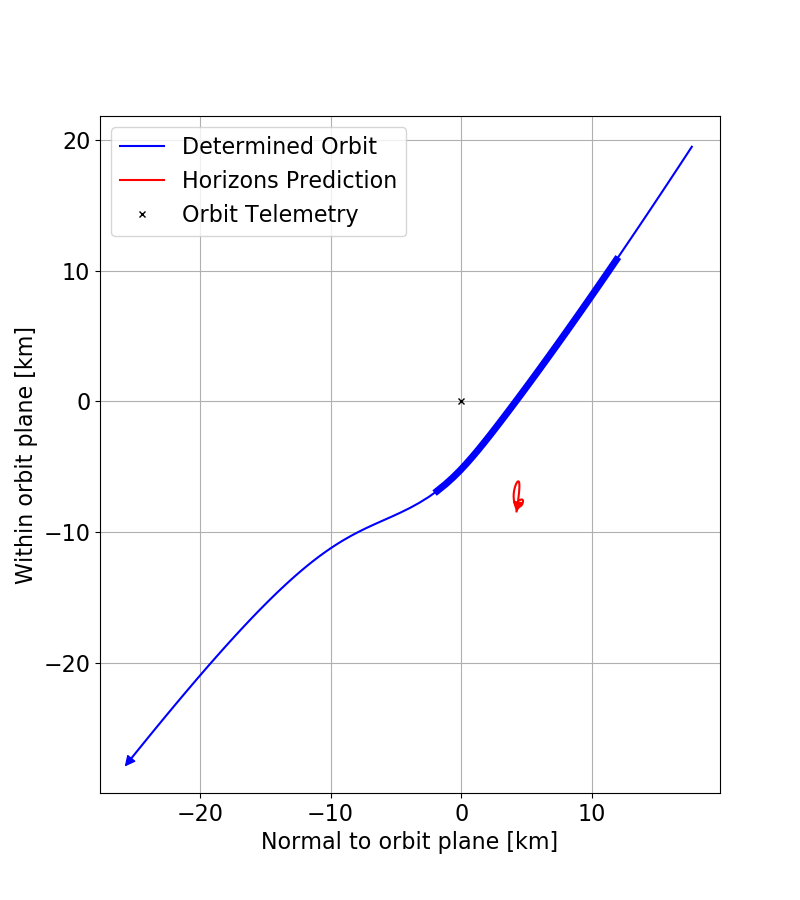}
    \caption{The subtle cross-track differences of the Horizons predicted trajectory and the determined orbit to the telemetry trajectory. The z-axis (into the page) and the y-axis correspond to the direction of the telemetry's velocity vector and the direction of Earth, respectively. The observed section of the determined orbit is highlighted by a thicker blue line. For reference, the distance travelled along the z-axis of this plot is 109,262~km, further enforcing how similar these orbits are.}
    \label{fig:cross-tracks}
\end{figure}

As shown in Fig.~\ref{fig:orbits}, the three orbits are so close to one-another that they are practically indistinguishable at this scale. Likewise, comparing the numbers directly from Table~\ref{table:orbits} can be misleading and does not show the subtle differences between them. In order to visualise the subtle variation between these orbits, we investigate the cross-track differences to the telemetry by constructing a frame of reference centred around and travelling with the telemetry trajectory, as shown in Fig.~\ref{fig:cross-tracks}. This telemetry "body frame" approach highlights the slight orbit differences throughout the flyby trajectory, where points in the body's x-y frame (Fig.~\ref{fig:cross-tracks}) represent parallel trajectories.

Fig.~\ref{fig:cross-tracks} shows how parallel the Horizons predicted trajectory is to the orbital telemetry as indicated by the relatively small deviation over the 12 hour window. Additionally, we note how accurate the determined orbit is to the telemetry, with only about 4 hours of measurements using human pointed off-the-shelf cameras.

Interestingly, if we use the observation range to convert the angular cross-track residual standard deviation into an equivalent residual distance, we get a linear cross-track residual standard deviation of about 31.2 km and 6.2 km corresponding to the start and the end of the observation window respectively; well within one standard deviation of the telemetry orbit, i.e. the x-axis of Fig.~\ref{fig:cross-tracks}. 

\section{The Creation of FireOPAL}
The OSIRIS-REx observation campaign turned out to be an excellent proof-of-concept for a networked design for SSA. As demonstrated in Section~\ref{ssec:orbit_comparision}, we were able to successfully recreate a NEO-like orbit using a distributed network of relatively low-resolution off-the-shelf imagers. Additionally, this achievement proved the capability of the developed and automated data reduction pipeline. To expand this successful proof-of-concept into the observation of Earth-bound objects and space debris, the prototype hardware, observatory structure, and data pipeline were adjusted into a more durable SSA system now known as FireOPAL; a partnership project between Curtin University and Lockheed Martin, Australia.

\begin{figure}
    \centering
    \includegraphics[width=\textwidth]{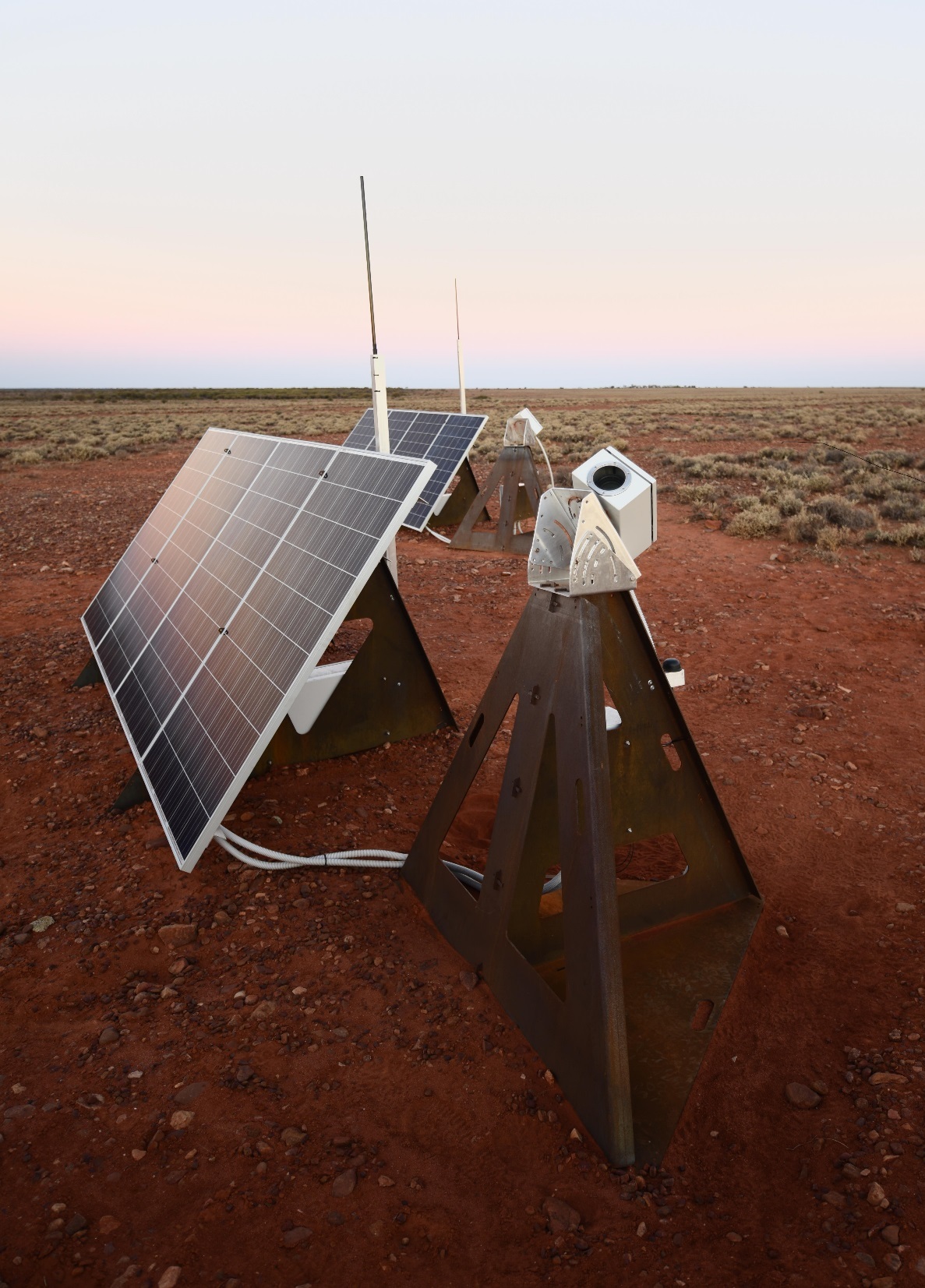}
    \caption{Two current FireOPAL units deployed at Glendambo that are part of six within South Australia. These six observatories form one node that tiles the GEO belt, with two additional similar nodes located in Western Australia and New South Wales.}
    \label{fig:glendambo_obs}
\end{figure}

Perhaps the biggest modification in the development of FireOPAL (Fireball OPtical ALert) was made to the observatory hardware. Each observatory is now completely autonomous and weather-hardened, consisting of a stand-alone and sturdy fixed-pointing sensor, four large solar panels, a set of high-capacity batteries, and an on-board computer to perform streak determination and light-curve extraction remotely and automatically in the field. The comparison of the 85\,mm and 105\,mm lenses during the OSIRIS-REx campaign ultimately influenced our choice to use the 105\,mm lens in the FireOPAL units due to its increased sensitivity. For reference, Fig.~\ref{fig:glendambo_obs} shows two FireOPAL observatories deployed in Glendambo, South Australia. 

These fixed-pointing sensors are arranged and aligned to maximise the number of recorded satellites from all orbit regimes, with overlap that enables triangulation for more accurate orbit determination. Currently, there are three clusters at different longitudes across Australia that each comprise of 6+ sensors tiling the GEO belt like a "fence". Sufficiently large satellites and space debris that pass through this field-of-view during the terminator window are observed by the network.

The OSIRIS-REx streak detection and light curve extraction algorithms have been adjusted to minimise noise-level artefacts while maintaining the majority of true-positive detections. A complete account of the FireOPAL image processing pipeline is detailed by \citet{madsen_fireopal:_2018}. Additionally, the orbit determination algorithm used in the calculation of OSIRIS-REx's slingshot trajectory has also evolved into a fully autonomous data association and orbit determination/refinement pipeline. Ambiguous streak directions are now automatically determined within the data association modules. Furthermore, the hyperbolic orbit propagation module was swapped out for the high-fidelity OreKit propagator \citep{maisonobe_orekit:_2010} to improve the accuracy of the estimated satellite orbits.

In late 2018, \citet{madsen_fireopal:_2018} revealed the first results of the FireOPAL network. These systems were shown to record about 4,000 images a night, detecting around 1,500 LEO streaks and 1,000 GEO measurements. When comparing the GEO measurements from one clear night to known objects in the SpaceTrack catalogue, FireOPAL was found to have a 90\% detection rate of targets within an observatory's field-of-view.

\citet{madsen_fireopal:_2018} also reported successful collaborative hand-off experiments of LEO and GEO satellites to a narrow field-of-view telescope operated by the Australian Defence Science \& Technology Group (DSTG) 24 hours after their latest viewing. This hand-off relied on the fully autonomous orbit determination and refinement pipeline to both construct the orbits and predict the future locations of a number of satellite objects. Over 90\% of the hand-off objects were detected by the DSTG sensor over a two week period in June 2018. For more FireOPAL performance and results, please refer to \citet{bland_fireopal:_2018} and \citet{madsen_fireopal:_2018}. 

At our current rate of data accumulation over the 21 nodes within FireOPAL network, we expect to record over 7 million LEO streaks and over 5 million GEO observations in one year. These numbers only set to increase in the future when more FireOPAL sensors are distributed at different longitudes around the world, leading to an SSA network capable of high satellite custody. This not only makes it easier to detect and follow satellites through orbital manoeuvres, but it aids greatly in collision avoidance and overall space traffic management in an effort to avoid the potential Kessler event \citep{kessler_collision_1978}.


\newpage
\section{Conclusion}
By using a distributed network of relatively low-resolution ground-based off-the-shelf sensors, we were able to successfully detect and determine the NEO-like orbit of NASA's OSIRIS-REx spacecraft on its scheduled slingshot trajectory around Earth. On the night of September 22nd, 2017, 13 teams from 6 Australian institutions spread across Australia and New Zealand to optimise triangulation capacity. In all, the observation campaign generated over 13,000 images within about four hours. 

As manual reduction of this dataset was unfeasible, an automated streak detection and orbit determination pipeline was developed. The streak detection algorithm found 1,045 faint OSIRIS-REx streaks, equating to 2,090 line-of-sight measurements; 99 of which were discarded as outliers. The recreated orbit of OSIRIS-REx was within about 10 km of orbit telemetry and closer than the predicted flyby orbit of a month before. 

Following the successful result of the OSIRIS-REx observation campaign, the FireOPAL project was initiated to transform this proof-of-concept into a more permanent SSA observation network. With this distributed network approach to SSA, we can now achieve both a high astrometric precision with a large field-of-view sensor, thereby removing the trade-off faced by traditional single-sensor designs.




\section{Acknowledgements}

This work was funded by the Australian Research Council as part of the Australian Discovery Project scheme, with funding from the Australian Government and the Government of Western Australia. This work was also supported by an Australian Government Research Training Program (RTP) Scholarship.

This research made use of Astropy, a community-developed core Python package for Astronomy \citep{collaboration_astropy:_2013}. Some figures in this work were generated using Matplotlib, another community-developed Python package \citep{hunter_matplotlib:_2007}.

\newpage
\section{References}
    \bibliography{research}{}
    \bibliographystyle{abbrvnat}

\end{document}